\documentclass[%
 reprint,
 amsmath,
 amssymb,
 aps,
 nofootinbib
]{revtex4-2}

\usepackage{physics}
\usepackage{graphicx}
\usepackage{dcolumn}
\usepackage{bm}

\usepackage[parfill]{parskip}

\usepackage[usenames,dvipsnames]{xcolor}
\usepackage{hyperref}
\hypersetup{
    pdfencoding=unicodge,
	colorlinks=true,
	urlcolor=Maroon,
	linkcolor=RoyalBlue,
	citecolor=Maroon,
	pdftitle={},
	pdfauthor={},
	pdfdisplaydoctitle=true,
	pdfstartview=FitH,
	linktocpage=true
}
\usepackage{amsmath}
\usepackage{dsfont}
\usepackage{amsfonts}
\usepackage{amssymb}
\usepackage{mathtools}
\usepackage{microtype}
\usepackage{bm}
\usepackage[utf8]{inputenc}
\usepackage{blkarray}
\usepackage{bigstrut}
\usepackage{nccmath}
\usepackage{enumitem}
\usepackage{float}
\usepackage{braket}
\usepackage{dsfont}
\usepackage[export]{adjustbox}
\usepackage[english]{babel}
\usepackage{stmaryrd}
\usepackage{orcidlink}
\usepackage{tikz}
\usepackage{bbm}
\usetikzlibrary{decorations.pathmorphing,decorations.pathreplacing,calligraphy,calc,snakes,cd,external,arrows.meta}
\tikzset{
    vector/.style={
        decoration={snake, aspect=0.75, mirror, segment length=2mm},
        decorate
    },
	photon/.style={decorate, decoration={snake, amplitude=1pt, segment length=6pt}
	}
}

\def\e{\mathrm{e}}
\def\D{\mathrm{D}}
\def\d{\mathrm{d}}

\def\mbs#1{\mathbf{#1}}

\def\>{\rangle}
\def\<{\langle}
\def\ot{\leftarrow}
\newcommand{\dbar}{{\text{d}\hspace*{-0.15em}\bar{}\hspace*{0.1em}}}
\DeclareFontFamily{U}{mathx}{}
\DeclareFontShape{U}{mathx}{m}{n}{<-> mathx10}{}
\DeclareSymbolFont{mathx}{U}{mathx}{m}{n}
\DeclareMathAccent{\widehat}{0}{mathx}{"70}
\DeclareMathAccent{\widecheck}{0}{mathx}{"71}
\usepackage{cleveref}
\newcommand{\myparagraph}[1]{\paragraph*{\textbf{\textbf{#1}}}}
 
\DeclareMathOperator*{\sumint}{%
\mathchoice%
  {\ooalign{$\displaystyle\sum$\cr\hidewidth$\displaystyle\int$\hidewidth\cr}}
  {\ooalign{\raisebox{.14\height}{\scalebox{.7}{$\textstyle\sum$}}\cr\hidewidth$\textstyle\int$\hidewidth\cr}}
  {\ooalign{\raisebox{.2\height}{\scalebox{.6}{$\scriptstyle\sum$}}\cr$\scriptstyle\int$\cr}}
  {\ooalign{\raisebox{.2\height}{\scalebox{.6}{$\scriptstyle\sum$}}\cr$\scriptstyle\int$\cr}}
}

\begin{document}

\title{Fourier Calculus from Intersection Theory}
\author{Giacomo Brunello$\,^{a,b,c}$}
\email{giacomo.brunello@phd.unipd.it}
\author{Giulio Crisanti$\,^{a,b}$}
\email{giulioeugenio.crisanti@phd.unipd.it}
\author{Mathieu Giroux$\,^{d}$}
\email{mathieu.giroux2@mail.mcgill.ca}
\author{Pierpaolo Mastrolia$\,^{a,b}$}
\email{pierpaolo.mastrolia@unipd.it}
\author{Sid Smith$\,^{a,e}$}
\email{sid.smith@ed.ac.uk}

\affiliation{$^a$Dipartimento di Fisica e Astronomia, Universita di Padova, Via Marzolo 8, 35131 Padova, Italy}
\affiliation{$^b$INFN, Sezione di Padova,
Via Marzolo 8, I-35131 Padova, Italy.}
\affiliation{$^c$Institut de Physique Théorique, CEA, CNRS, Université Paris-Saclay, F–91191 Gif-sur-Yvette cedex, France}
\affiliation{$^d$Department of Physics, McGill University, 3600 Rue University, Montr\'eal, H3A 2T8, QC Canada}
\affiliation{$^e$Higgs Centre for Theoretical Physics, University of Edinburgh, James Clerk Maxwell Building,Peter Guthrie Tait Road, Edinburgh, EH9 3FD, United Kingdom}

\begin{abstract}
Building on recent advances in studying the co-homological properties of Feynman integrals, we apply intersection theory to the computation of Fourier integrals. We discuss applications pertinent to gravitational bremsstrahlung and deep inelastic scattering in the saturation regime. 
After identifying the bases of master integrals, the latter are evaluated by means of the differential equation method. 
Finally, new results with exact dependence on the spacetime dimension $\mathrm{D}$ are presented.
\end{abstract}

\maketitle

\allowdisplaybreaks
\raggedbottom
%----------------------------------
%----------------------------------
\section{\label{sec:intro}Introduction}

In recent years, monumental efforts have been invested in developing tools for evaluating Feynman integrals in particle physics. 
Modern state-of-the-art computations face the challenge of applying integration-by-parts (IBP) decompositions \cite{Chetyrkin:1981qh, Laporta:2000dsw} in the most efficient way possible. IBP identities are relations among Feynman integrals sharing a common set of denominators, appearing with different propagator powers (irreducible scalar products in the numerator can be dealt with as denominators with negative powers). These allow for the decomposition of any Feynman integral in terms of a finite spanning set of simpler and linearly independent integrals, often referred to as \emph{master integrals}. The decomposition into master integrals is fundamentally a matter of linear algebra, and several publicly available computer programs can efficiently execute it \cite{Anastasiou:2004vj,vonmanteuffel2012reduze,Lee:2013mka,Georgoudis:2016wff,Smirnov:2019qkx,Klappert:2020nbg,Klappert:2019emp,Peraro:2019svx,Klappert:2020nbg,Wu:2023upw}.

When recognized as \emph{twisted periods}, Feynman integrals can alternatively be decomposed into master bases by making use of concepts and computing tools of \emph{de Rham twisted cohomology theory} \cite{cho1995,ojm/1200788347,Matsumoto1998-2,majima2000},
as first proposed in ~\cite{Mastrolia:2018uzb,Frellesvig:2019kgj,Frellesvig:2019uqt,Frellesvig:2020qot}.

Within this framework, integrals are considered 
as pairings of regulated integration domains and differential $n$-forms, known as
twisted cycles and cocycles respectively, 
which are elements of isomorphic vector spaces, equipped with inner products, called the \emph{intersection numbers}. 
The intersection number can be used to derive the decomposition of Feynman integrals in terms of master integrals by projections, as an alternative to the system-solving procedure underpinning IBP decompositions.

Although the most recent applications of intersection theory have dealt with Feynman integrals \cite{Frellesvig:2019kgj,Frellesvig:2019uqt,Mizera:2019vvs,Frellesvig:2020qot,Weinzierl:2020gda,Chen:2020uyk,Chen:2022lzr,Cacciatori:2021nli,Caron-Huot:2021iev,Caron-Huot:2021xqj,Giroux:2022wav,Chen:2023kgw,Fontana:2023amt,duhr2023feynman,Brunello:2023rpq}, the method is rather general, and the range of its applications can be extended to a much wider class of cases, relevant for physical and mathematical studies  
(see, e.g., \cite{cacciatori2022intersection,Gasparotto:2022mmp,Gasparotto:2023roh,de2023cosmology}). 

In this letter, we propose an intersection-theory based approach to the evaluation of dimensionally regularized Fourier transforms, herein referred to as \emph{Fourier integrals}.

Indeed, as is the case for any Feynman integral, expressing any Fourier integral in \emph{Baikov representation} \cite{Baikov:1996iu} enables its identification as a twisted period. Thus, relations between Fourier integrals and, in particular, their decompositions into a common master integral basis, can be obtained directly from intersection numbers. 
Moreover, intersection theory can be used to derive the system of differential equations satisfied by the master Fourier integrals, which can be solved analytically (when possible) 
similarly to Feynman integrals. \cite{Kotikov:1991pm,Remiddi:1997ny,Henn:2013pwa,Frellesvig:2017aai}. 

To illustrate our method, we identify three cases of physical interest, requiring the evaluation of certain types of dimensionally regulated Fourier integrals: \ref{ssec:a} the scalar (Feynman) propagator in position space, \ref{ssec:b} the tree-level gravitational spectral waveform, and \ref{ssec:c} color dipole scattering in high-energy QCD. In each of the considered cases, intersection numbers are used to build linear relations and differential equations for the associated master integrals in Baikov representation. Once the systems of differential equations are obtained, the solutions are systematically derived, provided an appropriate set of boundary conditions. Using this approach, we present \emph{new}, closed-form formulae for $\D$-dimensional Fourier integrals relevant to cases \ref{ssec:b} and \ref{ssec:c}. 

From a mathematical point of view, 
our results offer a generalization of the studies carried out on confluent hypergeometric integrals \cite{Matsumoto1998-2,majima2000}, involving intersection numbers between $1$-forms, to cases where the evaluation of intersection numbers for $n$-forms is required.

This letter is organized as follows. In sec.~\ref{sec:background} we provide the necessary background on intersection theory and outline its application to Fourier integrals. The method is then applied in sec.~\ref{sec:examples}, using the three cases of study (\ref{ssec:a}, \ref{ssec:b} and \ref{ssec:c}) mentioned above. For \ref{ssec:b} and \ref{ssec:c}, a minimal physics background is provided for the reader's convenience. In sec.~\ref{sec:conclusion}, we present our conclusions and an outlook. The four appendices contain the derivation of the Baikov representation for Fourier integrals, and auxiliary formulae recalled in the text.

%-------------------------------------------------------------------------------------------
\section{\label{sec:background}Fourier integrals and intersection theory}
In this section, we provide some background material on intersection theory and describe how it can be applied to dimensionally regularized Fourier integrals \cite{Mastrolia:2018uzb,Frellesvig:2019kgj,Frellesvig:2019uqt,Frellesvig:2020qot}
.\\
%----------------------------------
\myparagraph{Twisted cohomology}
We consider instances of \emph{twisted period integrals}, which generically take the form
\begin{align}
    I = \int_{C_{R}}u(\mbs{z})\varphi_{L}(\mbs{z})\,,\label{eq:twisted}
\end{align}
where the \emph{twist} $u(\mbs{z})$ is a multivalued function, $\varphi_{L}(\mbs{z})$ is an algebraic differential $n$-form and $C_{R}$ is a contour of integration.\footnote{It is assumed that the causality conditions have already been incorporated into the definition of $C_R$, which means the $i\varepsilon$ prescription is accounted for there, and not in the integrand.} The latter is defined such that $u(\mbs{z})$ vanishes on its boundary: $u(\mbs{z})=0$ for any $\mbs{z}\in \partial C_{R}$. This condition on $u$ together with eq.~\eqref{eq:twisted} give an equivalence relation between differential forms
\begin{align}
    \varphi_{L} \sim \varphi_{L}+\nabla_{\omega}\xi\,,\label{eq:eqClass}
\end{align}
where $\xi$ is a differential $(n{-}1)$-form and
\begin{equation}
    \nabla_{\omega}=\d+\omega\wedge\quad \text{with} \quad \omega=\d\log{u}\,.
\end{equation}

The collection of all the equivalence classes forms the \emph{twisted cohomology group} $H_{\omega}^{n}$.\footnote{Setting $\omega=0$ in eq.~\eqref{eq:eqClass}, one obtains the standard (non-twisted) de Rham equivalence class.} This group is always finite-dimensional and forms a vector space \cite{Frellesvig:2019kgj}. We denote an element by
\begin{align}
    \bra{\varphi_{L}}\in H_{\omega}^{n}\,.
\end{align}
As for any other finite dimensional vector space, there exists a \emph{dual} vector space $H_{-\omega}^{n}$, denoted by
\begin{align}
    H_{-\omega}^{n}\ni \ket{\varphi_{R}}\,.
\end{align}
Using these definitions, we write the (dual) integrals in eq.~\eqref{eq:twisted} as a \emph{pairing} between a (dual) cycle and a (dual) cocycle
\begin{subequations}
    \begin{align}
    I &= \int_{C_{R}}u(\mbs{z})\varphi_{L}(\mbs{z}) = \bra{\varphi_{L}}C_{R}]\,,\\
    \widecheck{I} &= \int_{C_{L}}u(\mbs{z})^{-1}\varphi_{R}(\mbs{z})= [C_{L}\ket{\varphi_{R}}\,.
\end{align}
\end{subequations}

%----------------------------------
\myparagraph{Basis of master integrals}

We can formally define bases for $H_{\omega}^{n}$ and its dual $H_{-\omega}^{n}$ as
\begin{subequations}
    \begin{align}
    \text{span}\{\bra{e_{1}},\ldots,\bra{e_{\nu}}\}&=H_{\omega}^{n}\,,\\
     \text{span}\{\ket{\widecheck{e}_{1}},\ldots,\ket{\widecheck{e}_{\nu}}\}&= H_{-\omega}^{n}\,,
\end{align}
\end{subequations}
respectively. Both $H_{\omega}^{n}$ and $H_{-\omega}^{n}$ have the same dimension \cite{Lee:2013hzt,Frellesvig:2019uqt}, which can be computed as
\begin{align}\label{eq:noCP}
    \nu = \left\{\#~\text{of solutions to $\omega=0$}\right\}\,.
\end{align}
Any elements $\bra{\varphi_{L}}\in H_{\omega}^{n}$ and $\ket{\varphi_{R}}\in H_{-\omega}^{n}$ can then be decomposed with respect to the choice of (dual) bases
\begin{equation}\label{eq:masterDec}
    \bra{\varphi_{L}} = \sum_{i=1}^{\nu}c_{i}\bra{e_{i}}
    \quad\text{and} \quad
      \ket{\varphi_{R}} = \sum_{i=1}^{\nu}\widecheck{c}_{i}\ket{\widecheck{e}_{i}}\, .
\end{equation}
By pairing the linear combinations in eq.~\eqref{eq:masterDec} with the appropriate contours, we obtain the decompositions of the corresponding twisted periods with respect to the basis of master integrals $J_{i}=\bra{e_{i}}C_{R}]$ and $\widecheck{J}_{i}=[C_{L}\ket{\widecheck{e}_{i}}$
\begin{subequations}
    \begin{align}
    I &= \bra{\varphi_{L}}C_{R}] = \sum_{i=1}^{\nu}c_{i}\bra{e_{i}}C_{R}] = \sum_{i=1}^{\nu}c_{i}J_{i}\,,\\
    \widecheck{I} &= [C_{L}\ket{\varphi_{R}} = \sum_{i=1}^{\nu}\widecheck{c}_{i}[C_{L}\ket{\widecheck{e}_{i}} = \sum_{i=1}^{\nu}\widecheck{c}_{i}\widecheck{J}_{i}\,.
\end{align}
\end{subequations}

%----------------------------------
\myparagraph{Integral decompositions}

Following \cite{ojm/1200788347,Mastrolia:2018uzb}, we can introduce a scalar product $\braket{\bullet,\bullet}$ between elements of $H_{\omega}^{n}$ and $H_{-\omega}^{n}$, called the \emph{intersection number}. With this additional structure, the coefficients $c_i$ can be extracted using the \emph{master decomposition formula} \cite{Mastrolia:2018uzb,Frellesvig:2019kgj}
\begin{equation}
    c_{i} = \sum_{j=1}^{\nu}\braket{\varphi_{L},\widecheck{e}_{j}}(C^{-1})_{ji}\quad \text{with} \quad C_{ij} = \braket{e_{i},\widecheck{e}_{j}}\,.\label{eq:coef0} 
\end{equation}
The computation of intersection numbers in eq.~\eqref{eq:coef0} has been the primary focus of recent work in intersection theory, with significant progress made over the past few years \cite{Weinzierl:2020xyy,Frellesvig:2020qot,Frellesvig:2021vem,Mandal,Chestnov:2022xsy,Fontana:2023amt}.

In the univariate case ($n=1$), a compact formula for the intersection number is known and given by
\begin{align}
    \braket{\varphi_{L},\varphi_{R}} = \sum_{p\in\mathcal{P}}\text{Res}_{z=p}\left(\psi\varphi_{R}\right)\,,
\end{align}
where $\mathcal{P}$ is the set of $\omega$'s poles, and $\psi$ satisfies the differential equation
\begin{align}
    \nabla_{\omega}\psi=\varphi_{L}\,.
\end{align}
The calculation of multivariate intersection numbers requires more effort and there are several strategies \cite{ojm/1200788347,Mizera:2019gea,Frellesvig:2019uqt,Mizera:2019vvs,Caron-Huot:2021iev,Chestnov:2022alh,Chestnov:2022xsy,Giroux:2022wav,Fontana:2023amt}. In this letter, we adopt the ones introduced in \cite{Fontana:2023amt,Caron-Huot:2021iev,Caron-Huot:2021xqj,Frellesvig:2019uqt,Brunello:2023rpq}.

The computation of intersection numbers allows us to build the \emph{differential equation} $\boldsymbol{\Omega}_{x}$ satisfied by the basis of master integrals in any external variable $x$
\begin{align}
\partial_{x}J_{i} = [\boldsymbol{\Omega}_{x}]_{ij}J_{j}\,.\label{eq:de0}
\end{align}
To see this, we note that in the language of twisted cohomology, eq.~\eqref{eq:de0} translates to
    \begin{align}
    \partial_{x}\bra{e_{i}} &= \bra{\partial_{x}(ue_{i})/u} =  [\boldsymbol{\Omega}_{x}]_{ij}\bra{e_{j}}\,,
\end{align}
which implies
\begin{align}\label{eq:deInt}
    [\boldsymbol{\Omega}_{x}]_{ij} &= \braket{\partial_{x}(ue_{i})/u,\widecheck{e}_{k}}[C^{-1}]_{kj}\,.
\end{align}
We reiterate that the derivations of \cref{eq:coef0,eq:deInt} do not involve solving (potentially large) systems of linear equations but instead exclusively rely on the computation of intersection numbers.\\

%----------------------------------
\myparagraph{Fourier integrals in Baikov representation}
We consider a generic $\D$-dimensional \emph{Fourier integral}, which takes the form
\begin{subequations}\label{eq:genericFourier}
\begin{align}
    &\tilde{f}(\{x_{i}\})=\int f(\{q_{i}\})\prod_{j=1}^{L}\e^{iq_{j}\cdot x_{j}}\dbar^\D q_j\,,
    \\& \text{with measure:}\qquad \dbar^{\D} q_j=\frac{\d^{\D}q_{j}}{(2\pi)^{\D/2}}\,.
\end{align}
\end{subequations}
Eq.~\eqref{eq:genericFourier} is the Fourier transform of the function/distribution $f$ performed over $L$ internal vectors $\{q_{i}\}$. The result is a function of $E\geq L$ external vectors $\{x_i\}$. We denote the set of $n= \frac{L}{2}(L{+}1){+}LE$ internal scalar products as
\begin{align}
    \hspace{-0.2cm} S = \{q_{1}^{2},q_{1}\cdot q_{2},\dots,q_{L}^{2},q_{1}\cdot x_{1},q_{1}\cdot x_{2},\dots,q_{L}\cdot x_{E}\}\,.
\end{align}
To reinterpret the Fourier transform in eq.~\eqref{eq:genericFourier} as a twisted period, we propose to change variables to the \emph{Baikov variables} \cite{Baikov:1996iu,Grozin}: the procedure involves a first change of variables from the internal vectors $q_{i}$ to the internal scalar products, followed by a second change of variables,
\begin{align}
    z_{i} = A_{ij} \, S_{j}+B_{j}\,,
\end{align}
where $A$ is an $n\times n$ matrix, $B$ is an $n$-dimensional vector and $S_j$ is the $j^\text{th}$ element of $S$. Both operations only depend on the external scalar products. Once the dust settles, the result reads
\begin{align}\label{eq:FourierBaikov}
    \tilde{f} = \int_{C_{R}}u(\mbs{z}) \, \varphi_{L}(\mbs{z})\,,
\end{align}
where
\begin{equation}
    C_{R}=\bigcap_{i=1}^{L}\left\{ \frac{\det G_{\{q_i, \ldots, q_L, x_1, \ldots, x_E\}} }{\det G_{\{q_{i+1}, \ldots, q_L, x_1, \ldots, x_E\}}} > 0 \right\}\,,
\end{equation}
is the contour of integration. 
The differential form 
$\varphi_{L}(\mbs{z}) = f(\mbs{z}) \, \d^{n}\mbs{z}$ contains the function/distribution $f$ we would like to Fourier transform and
\begin{align}
    u(\mbs{z}) = \kappa\ \e^{ig(\mbs{z})}B(\mbs{z})^{\frac{\D-L-E-1}{2}}\,,
\end{align}
is the \emph{twist}. Here, $g(\mbs{z})$ is always \emph{linear} in  $\mbs{z}$ and we define 
\begin{subequations}
    \begin{align}
    G(\{x\}) &= \det[x_{i}\cdot x_{j}]\,,\\
    B(\mbs{z}) &= G(q_{1},\dots,q_{L},x_{1},\dots,x_{E})\,,\label{eq:baikovPoly}\\
    \kappa&=\frac{\pi^{\frac{L(1-L-2E)}{4}}G(x_{1},\dots,x_{E})^{\frac{E-\D+1}{2}}}{2^{\frac{L\D}{2}}\det A\prod_{j=1}^{L}\Gamma\left(\frac{\D{-}L{-}E+j}{2}\right)}\,.\label{eq:baikovconst}
\end{align}
\end{subequations}

Note that $B\geq0$ on $C_R$. Complementary details regarding the derivation of eq.~\eqref{eq:FourierBaikov} can be found in app.~\ref{sec:BaikovFourier}.

Representing a Fourier integral as the twisted period in eq.~\eqref{eq:FourierBaikov} enables the use of intersection theory for the construction of differential equation (c.f., eq.~\eqref{eq:de0}). Thus, the master Fourier integrals $J_i$ can be evaluated by solving the system of differential equations, analogously to Feynman integrals. 

%----------------------------------
%----------------------------------

\section{\label{sec:examples}Applications}

In this section, we apply the formalism described above to three families of Fourier integrals arising in various corners of particle physics. An ancillary \textsc{Mathematica} file (\texttt{ancillary.m}) containing complementary details for each example is given as supplementary material. 

Below, $\mathcal{M}=\mathbbm{R}^{1,\D-1}$ denotes the Minkowski spacetime manifold. Unless specified otherwise, we work in the mostly plus Lorentzian signature $(-,+,+,...,+)$.
%
%----------------------------------
\subsection{Fourier transform of a scalar propagator}\makeatletter\def\@currentlabel{(A)}\makeatother\label{ssec:a}
As a first example, we consider the Fourier transform of a massive scalar Feynman propagator,
\begin{align}
    I_{n} = \int_{\mathcal{M}}\dbar^{\D} q\frac{\e^{iq\cdot x}}{(q^{2}+m^{2}-i\varepsilon)^{n}}\,.\label{integral1}
\end{align}
 We work with dimensionless integrals $K_n$, defined by
\begin{align}\label{eq:Kndef}
    I_{n} = m^{\D{-}2n}K_{n}\,, ~ ~ K_{n} = \int_{\mathcal{M}} \dbar^{\D}k\frac{\e^{ik\cdot v}}{(k^{2}{+}1{-}i\varepsilon)^{n}}\,,
\end{align}
where $v=mx$ and $k=q/m$ are both dimensionless vectors.
For $K_n$, we have $L=1$ internal vector $\{k\}$ and $E=1$ external vector $\{v\}$. We define the $n=2$ integration variables as $z_{1}=k^{2}+1$ and $z_{2}=k\cdot v$. Thus, in the Baikov representation, this integral takes the form
\begin{align}\label{eq:Kdef}
    K_{n} = \int\frac{\d^{2} \mbs{z}}{z_{1}^{n}}u(\mbs{z}) \ , 
\end{align}
where the twist is given by
\begin{equation}
    u(z_{1},z_{2}) =  \frac{\e^{iz_{2}}\tau^{\frac{2-\D}{2}}}{2^{\frac{\D}{2}}\sqrt{\pi}\Gamma((\D-1)/2)} ((z_{1}-1)\tau-z_{2}^2)^{\frac{\D-3}{2}}\,,
\end{equation}
and $\tau= v^2$. From eq.~\eqref{eq:noCP}, the number of master integrals is found to be $\nu=2$. We choose them as $K_{1}$ and $K_{2}$ and form the basis vector $\mbs{K}=(K_{1},K_{2})^\top$.
From intersection decompositions, we find that $\mbs{K}$ obeys the differential equation
\begin{align}\label{eq:system}
    \partial_{\tau}\mbs{K}&= \boldsymbol{\Omega}_{\tau}\cdot\mbs{K} 
    \ , \quad {\rm with} \quad 
    \boldsymbol{\Omega}_{\tau} = -\begin{pmatrix}
        \frac{(\D-2)}{2\tau}&\frac{1}{\tau}\\
        \frac{1}{4}&0
    \end{pmatrix}\,.
\end{align}
We can decouple this system of equations into second- and first-order differential equations for $K_{1}$ and $K_{2}$ respectively, by first changing the variable $\tau$ in favor of the intermediate variable $t=\sqrt{-\tau}$. We find
\begin{subequations}
    \begin{align}
          0&=t\partial^2_tK_1(t)+(\D-1)\partial_t K_1(t)+tK_1(t)\,,\label{eq:pde1}\\
          0&=K_2(t)+\frac{1}{2}t\partial_tK_1(t)+\frac{\D-2}{2}K_1(t)\,.\label{eq:pde2}
    \end{align}
\end{subequations}
Solving eq.~\eqref{eq:pde1} first, followed by eq.~\eqref{eq:pde2}, results in (after changing the variable back to $\tau$)
\begin{subequations}\label{eq:genK1sol}
\begin{align}
K_1&=\frac{c_1 J_{\frac{\D-2}{2}}\left(\sqrt{-\tau }\right)+c_2 Y_{\frac{\D-2}{2}}\left(\sqrt{-\tau }\right)}{(-\tau )^{\frac{\D-2}{4}} }\,,\\
K_2&=-\frac{c_1 J_{\frac{\D-4}{2}}\left(\sqrt{-\tau }\right)+c_2 Y_{\frac{\D-4}{2}}\left(\sqrt{-\tau }\right)}{2(-\tau )^{\frac{\D-4}{4}}}\,,
\end{align}
\end{subequations}
where $J_{\nu}(z)$ and $Y_{\nu}(z)$ are the Bessel functions of the first and second kind respectively.

To fix the boundary constants $c_1$ and $c_2$ for $\mbs{K}$, we consider its boundary values at $\tau=\infty$ and $\tau=0$. 

In terms of the four-vector $v$, the boundary point at $\tau=\infty$ is approached with constant and finite spacelike component $\mbs{v}$ and piecewise timelike component 
\begin{equation}\label{eq:v0}
    v^0\to\begin{cases}
        -i\infty & \text{for} \, k^0\ge0\,,\\
        i\infty & \text{for} \, k^0<0\,.
    \end{cases}
\end{equation}
In this limit, the integrand in eq.~\eqref{eq:Kndef} is exponentially suppressed
\begin{equation}\label{eq:expSupp}
\begin{split}
        \e^{ik\cdot v}&=\e^{i(-k_0 v^0+\mbs{k}\cdot\mbs{v})}\\&\sim\e^{-k_0 \vert v^0\vert }\Theta(k_0)+\e^{k_0 \vert v^0\vert}\Theta(-k_0) \quad (\vert v^0\vert\gg 1)\,,
\end{split}
\end{equation}
where $\Theta(z)$ denotes the Heaviside step function. Thus, $K_n$ vanishes in this limit for any $n\in\mathbbm{Z}$. At the level of eq.~\eqref{eq:genK1sol}, this condition enforces
\begin{equation}
    c_2=i\,c_1\,.
\end{equation}
Within dimensional regularization, the integral in eq. \eqref{eq:Kndef} can also be explicitly evaluated at $\tau=0$ (approached with $v^\mu=0$). To do so, one can first perform a Wick rotation on $K_n(v^\mu=0)$ and then use \cite[eq.(7.85)]{Peskin:1995ev} to find
\begin{equation}
    \begin{split}
         \int_{\mathcal{M}}\dbar^{\D} k\frac{1}{(k^{2}{+}1{-}i\varepsilon)^{n}}&=\frac{i\, \Gamma \left(n-\frac{\D}{2}\right)}{2^{\frac{\D}{2}}\Gamma (n)}\,.\label{known2}
    \end{split}
\end{equation}
When comparing the result of this calculation at $n=1$ or $n=2$ with eq.~\eqref{eq:genK1sol} in the vicinity of $\tau=0$, one finds that
\begin{equation}\label{eq:fixc10}
    c_1=\frac{\pi}{2}\e^{i \pi \D/2} \qquad (\D<2)\,.
\end{equation}
Putting everything together and substituting back $\tau=v^2=m^2 x^2$, we find
\begin{subequations}
    \begin{align}
\hspace{-0.2cm}I_1&=\frac{\pi  \e^{\frac{i \pi  \D}{2}}}{2} \left(-\frac{x^2}{m^2}\right)^{\frac{2-\D}{4}} H_{\frac{\D-2}{2}}^{(1)}\left(m \sqrt{-x^2}\right)\,,\label{eq:I1ans}\\
    \hspace{-0.1cm}I_2&=-\frac{\pi  \e^{\frac{i \pi  \D}{2}}}{4}  \left(-\frac{x^2}{m^2}\right)^{\frac{4-\D}{4}} H_{\frac{\D-4}{2}}^{(1)}\left(m \sqrt{-x^2}\right)\,,
\end{align}
\end{subequations}
where $H^{(1)}_{\nu}(z)=J_{\nu}(z)+iY_{\nu}(z)$ is the Hankel function of the first kind.\footnote{We note that since our original integral is manifestly spherically symmetric after performing Wick rotation, it is not surprising that its closed form involves Bessel functions, independently of spacetime dimension.}

Despite the dimensionality condition in eq.~\eqref{eq:fixc10}, we find the $\D\to4$ limit of eq.~\eqref{eq:I1ans} to be smooth, and for timelike, spacelike and lightlike $x$, respectively, to reduce to the known results in \cite[eqs. (2.84), (2.85) and (2.86)]{scalar}, once factors coming from different normalisations and metric signatures are taken into account (see also \cite{Cacciatori:2023tzp}). 

Moreover, taking the $m\to0$ limit, we also verify that $I_1$ and $I_2$ agree with the result for the massless integral
\begin{equation}\label{eq:masslesssRes}
   \int_{\mathcal{M}}\dbar^{\D}q\frac{\e^{iq\cdot x}}{(q^{2}{-}i\varepsilon)^{n}}=\frac{i\,\left(x^2\right)^{n-\frac{\D}{2}} \Gamma \left(\frac{\D}{2}{-}n\right)}{2^{\frac{4 n-\D}{2}}\Gamma (n)}\,,
\end{equation}
which can be obtained directly from the result in app.~\ref{sec:fourierPL}, after Wick rotating the left-hand side of eq.~\eqref{eq:masslesssRes}.

%----------------------------------
\subsection{Spectral gravitational waveform}\makeatletter\def\@currentlabel{(B)}\makeatother\label{ssec:b}
With the prospect of future space-based gravitational wave observatories such as LISA \cite{LISA:2022yao}, a pressing theoretical problem is to streamline the path to higher-precision computations as much as possible. There is a growing interest in obtaining, from amplitude methods, the one-loop correction to the gravitational waveform \cite{Cristofoli:2021vyo,Damgaard:2023vnx,Herderschee:2023fxh,Brandhuber:2023hhy,Kosower:2018adc,Georgoudis:2023lgf,caronhuot2023measured,bini2023comparing}, as well as the spin corrections \cite{DeAngelis:2023lvf,Brandhuber:2023hhl,Aoude:2023dui}  which could potentially be measured by such observatories.

\begin{figure}
    \centering
   \includegraphics[scale=1]{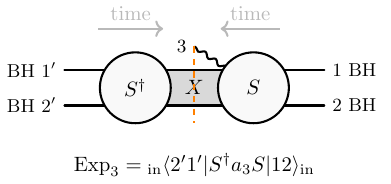}
    \caption{The KMOC momentum space waveform as the expectation value of measuring a graviton (labelled by $3$) in the background of two Schwarzschild black holes/heavy scalars (labelled by $1$ and $2$). The prime decorations on $1'$ and $2'$ emphasize that the scattering is non forward (no in and out states are exactly collinear). To obtain, e.g., the spectral waveform, one needs to Fourier transform this observable to impact parameter.}
    \label{fig:momentumSpaceWaveForm}
\end{figure}

A standard dictionary used to link waveforms and scattering amplitudes is the observable-based Kosower--Maybee--O’Connell (KMOC) formalism \cite{Kosower:2018adc}. In this framework, the gravitational waveform is characterized as the \emph{expectation value} $\text{Exp}_3$ of measuring asymptotically in the future a graviton field in the background of two black holes (modeled here as heavy scalars) scattering off each other, from and back to the far past (see Fig.~\ref{fig:momentumSpaceWaveForm}). 

To establish the connection with scattering amplitudes more precisely, it is useful to first introduce the generators $a, a^\dag, b$ and $b^\dag$ for the algebra of asymptotic measurements. Its existence is physically motivated by the naive expectation that finite energy excitations in the ``bulk'' should decay into a set of \emph{stable} and \emph{free} particles at asymptotic times. This means that the asymptotic states are assumed to be free of any external forces/fields, so that they do not radiate nor decay.\footnote{In the context of collider physics, particles encountered near the detectors are, of course, generally \emph{not} free (their motion is most likely affected by background fields). In such cases, it is essential to also consider the scattering of unstable particles (which can decay and radiate). Recent literature on this subtle subject includes \cite{Hannesdottir:2019umk,Hannesdottir:2019opa,Hannesdottir:2022bmo}.}

The annihilation and creation operators in the far past are denoted, respectively, by $a$ and $a^\dagger$, while those in the far future are similarly denoted by $b$ and $b^\dagger$. In what follows, the key property is that $a$ and $b$ are conjugated to each other with respect to unitary time evolution : $b=S^\dagger a S$ (and, similarly, $b^\dagger=S^\dagger a^\dagger S$), where $SS^\dag=\mathbbm{1}$.  We refer the reader to \cite{caronhuot2023measured} for complementary details.

The background in which the scattering occurs is defined by perturbations of the time-invariant vacuum $\vert0\>$ in the far past

\begin{equation}
    \vert 12\>=a_2^\dagger a_1^\dagger\vert 0\> \quad\text{and}\quad \vert 1'2'\>=a_{2'}^\dagger a_{1'}^\dagger\vert 0\>\,.
\end{equation}

As these two-particle states evolve over time, they can interact non-trivially with each other (i.e., create and absorb particles). Then, $\text{Exp}_3$ is defined as

\begin{equation}\label{eq:exp}
    \text{Exp}_3= {}_{\text{in}}\<2'1'\vert b_3\vert 12\>_{\text{in}}\,.
\end{equation}

The connection between $\text{Exp}_3$ and amplitudes is made manifest in two steps. First, using the relation $b=S^\dagger a S$ and inserting a complete basis of states\footnote{The symbol $\sumint_X$ formally denotes an integral-sum over the on-shell phase space of the inserted state $\ket{X}$ (see \cite[eq.~(3.5)]{caronhuot2023measured}).} $\mathbbm{1}=\sumint_X\vert X\>\< X\vert$ in eq.~\eqref{eq:exp}, we obtain

\begin{equation}\label{eq:exp2}
    \begin{split}
        \text{Exp}_3&={}_{\text{in}}\<2'1'\vert S^\dagger a_3 S\vert 12\>_{\text{in}}\\&=\sumint_X{}_{\text{in}}\<2'1'\vert S^\dagger\vert X\>\< X3\vert S\vert 12\>_{\text{in}}\,.
    \end{split}
\end{equation}

Next, plugging the decomposition formula $S=\mathbbm{1}+iT$ of the 4-point S-matrix (where $T$ is the connected part) into eq.~\eqref{eq:exp2}, we obtain

\begin{equation}
    \begin{split}
        \text{Exp}_3&={}_{\text{in}}\<32'1'\vert iT\vert 12\>_{\text{in}}\\&\qquad+\sumint_X{}_{\text{in}}\<2'1'\vert T^\dagger\vert X\>\< X3\vert T\vert 12\>_{\text{in}}\,.
    \end{split}
\end{equation}

The first term is a (conventional) time-ordered $3\ot 2$ amplitude, while the second term is a product of two time-ordered amplitudes glued together by a $s_{1'2'}=-(p_{1'}{+}p_{2'})^2$ channel cut. In practice, we can therefore compute $\text{Exp}_3$ perturbatively,  directly from conventional time-ordered Feynman rules. (Alternatively, it was recently explained in \cite{caronhuot2023crossing} how to obtain such observables from analytic continuations of time-ordered scattering amplitudes.)

To eventually streamline comparison with experimental data, one may opt to work with waveforms expressed as functions of variables other than momenta. Such quantities can be derived from $\text{Exp}_3$ after performing additional Fourier transforms. For example, obtaining the \emph{spectral} waveform requires to Fourier transform $\text{Exp}_3$ to impact parameter space. Similarly, to obtain the \emph{time domain} waveform, an additional Fourier transform in the frequency of the outgoing graviton is needed.

It was recently demonstrated in \cite{Herderschee:2023fxh} (see also \cite{Cristofoli:2021vyo}) that to obtain the leading-order (tree-level) spectral waveform in pure general relativity and $\mathcal{N}=8$ supergravity from KMOC formalism, (similar integrals appear also in the EFT approach \cite{Mougiakakos:2021ckm,Riva:2021vnj}, in the context of worldline QFT \cite{Jakobsen:2021smu,Jakobsen:2021lvp} and in the eikonal approach \cite{DiVecchia:2023frv}) one must perform Fourier transforms of the form\footnote{Note that the exponential has the non-standard sign. This is due to our use of a signature convention opposite to that in \cite{Herderschee:2023fxh,Cristofoli:2021vyo}.}

\begin{equation}\label{eq:ExpTreeGrav0}
    \begin{split}
\hspace{-0.33cm}\mathcal{I}_{\beta_1\beta_2}^{\boldsymbol{\nu}_{2m}}&=\int_{\mathcal{M}} \dbar^\D q\frac{\delta(u_1{\cdot}q)\delta(u_2{\cdot} ( q{-}k)) q^{\nu_1}{\dots}q^{\nu_{2m}}\e^{-i  q\cdot b}}{[ q^2-i\varepsilon]^{\beta_1}[( q{-}k)^2-i\varepsilon]^{\beta_2}}\,,
    \end{split}
\end{equation}

where the $u_i$\,s denote the (dimensionless) classical velocities of the heavy external objects, $k$ is the (on-shell: $k^2=0$) graviton momentum and $b$ the impact parameter.

In what follows, we showcase how our method can be applied to obtain new $\D$-dimensional closed form formulae for eq.~\eqref{eq:ExpTreeGrav0} in the case where $\beta_2=0$. (Given that the ultra-soft graviton limits of the results presented below are not smooth, a separate evaluation of eq.~\eqref{eq:ExpTreeGrav0} for $k^\mu=0$ is provided for completeness in app.~\ref{sec:exBeasy}.)\\

\myparagraph{Evaluation of $\mathcal{I}_{\beta_10}^{\boldsymbol{\nu}_{2m}}$}
The tensor reduction of eq.~\eqref{eq:ExpTreeGrav0} is simple to perform. We find

\begin{equation}\label{eq:ExpTreeGrav}
    \begin{split}
\mathcal{I}_{\beta_10}^{\boldsymbol{\nu}_{2m}}&=\mathcal{N}~\mathcal{I}_{\beta_1-m}\,,
    \end{split}
\end{equation}
where we define

\begin{subequations}
    \begin{align}
\mathcal{I}_{\alpha}&=\int_{\mathcal{M}}\dbar^\D  q \frac{\delta(u_1{\cdot}q)\delta(u_2{\cdot} ( q{-}k))\e^{-i  q\cdot b}}{[q^2-i\varepsilon]^{\alpha}
}
\,,\label{eq:scalarIntWaveform}\\    \mathcal{N}&=\frac{(-1)^m\big[\eta^{\otimes m}\big]^{\{\boldsymbol{\nu}_{2m}\}}}{\prod_{j=1}^{m}(\D{+}2(j{-}1))}\,.
    \end{align}
\end{subequations}
Above, $\eta^{\otimes m}$ denotes $m$ occurrences of the metric tensor, $[\dots]^{\{\boldsymbol{\nu}_{2m}\}}$ stands for the sum over all possible shuffling of the $2m$ Lorentz indices $\nu_i$ in the tensor $[\dots]$. 

Defining the kinematics as
\begin{equation}\label{eq:kinWF}
    \begin{aligned}
   u_i^2&=-1  \,,&    u_1\cdot u_2& =-y  \,, &  u_i \cdot b& = 0 \,,\\
      k \cdot b &= 0  \,,&  k\cdot u_1 &= -w_1  \,,&  k\cdot u_2 &= -w_2\,, 
\end{aligned}
\end{equation}
and denominators
\begin{equation}
      \hspace{-0.2cm} z_1 = u_1\cdot q\,, ~ z_2= u_2\cdot q+w_2\,, ~ z_3= q^2\,, ~ z_4= -i b\cdot q\,,
\end{equation}

the integral in Baikov representation reads: 
\begin{equation}
    \mathcal{I}_\alpha = \int_{\mathcal{M}}\frac{\d^{4}\mbs{z}}{z_3^\alpha} u(\mbs{z})\delta(z_1)\delta(z_2)\,.
\end{equation}
Treating the delta functions appearing as cut propagators, we can rewrite the integrals by taking the residues at $z_1=0$ and $z_2=0$ as
\begin{equation}
    \mathcal{I}_\alpha = \int_{\mathcal{M}}\frac{\d^{2}\mbs{z}}{z_3^\alpha} u(\mbs{z})\big\vert_{z_1=z_2=0}\,,
\end{equation}
where the twist on the cut is defined as
\begin{equation}
\begin{split}
        u(\mbs{z})\big\vert_{z_1=z_2=0}&=- i \e^{z_4}\frac{\left(-b^2 \left(y^2-1\right)\right)^{2-\frac{\D}{2}}}{2^{\frac{\D}{2}}\pi ^{3/2} \Gamma \left(\frac{\D-3}{2}\right)}\\
        &\times\left(b^2 w_2^2-\left(y^2-1\right) \left(b^2 z_3+z_{4}^2\right)\right)^{\frac{\D-5}{2}}\,.
\end{split}
\end{equation}

Rescaling with respect to $|b|$ allows us to work with dimensionless integrals $\mathcal{K}_n$ defined as
\begin{align}
    \mathcal{I}_{\alpha}=|b|^{2+2\alpha-\D}\mathcal{K}_\alpha \,.
\end{align} 
We find $\nu=2$ master integrals, which we define as $\boldsymbol{\mathcal{K}}=\{\mathcal{K}_1,\mathcal{K}_2\}$. 

Next, defining the dimensionless parameter $s=w_2^2\,b^2$, intersection decompositions yield the system of differential equations

\begin{subequations}
    \begin{align}\label{eq:system2}
    \partial_{s}\boldsymbol{\mathcal{K}}&= \boldsymbol{\Omega}_{s}\cdot\boldsymbol{\mathcal{K}}
    \ , \quad \quad 
    \boldsymbol{\Omega}_{s} = \begin{pmatrix}
        0&\frac{-1}{y^2-1}\\
        -\frac{1}{4 s}&\frac{\D-6}{2 s}
    \end{pmatrix}\,,\\
    \partial_{y}\boldsymbol{\mathcal{K}}&= \boldsymbol{\Omega}_{y}\cdot\boldsymbol{\mathcal{K}}
    \ , \quad \quad 
    \boldsymbol{\Omega}_{y} = \begin{pmatrix}
        \frac{-y}{y^2-1}&\frac{2 s y}{\left(y^2-1\right)^2}\\
        \frac{y}{2 \left(y^2-1\right)}&\frac{(5-\D) y}{y^2-1}
    \end{pmatrix}\,.
\end{align}
\end{subequations}
 
In terms of the dimensionful master integrals, the solution to eq.~\eqref{eq:system2} reads
\begin{align}
&\mathcal{I}_1=\left(b^2/w_2^2\right)^{\frac{4{-}\D}{4}} \left(y^2{-}1\right)^{\frac{2{-}\D}{4}}\label{eq:newExBMIsUnfixed}\\ 
&\times\left[c_1 I_{2-\frac{\D}{2}}\left(\frac{\sqrt{b^2} w_2}{\sqrt{y^2{-}1}}\right)-c_2 I_{\frac{\D}{2}-2}\left(\frac{\sqrt{b^2} w_2}{\sqrt{y^2{-}1}}\right)\right]\,,\notag
\end{align}

and similarly for $\mathcal{I}_2$. In this expression, $I_{\nu}(z)$ denotes the modified Bessel function of the first kind. 

We now fix the boundary conditions. Similarly to what happened in eq.~\eqref{eq:v0}, in the neighborhood of $b^2=\infty$ approached from
\begin{equation}\label{eq:b0}
    b^0\to\begin{cases}
        i\infty & \text{for} \, q^0\ge0\,,\\
        -i\infty & \text{for} \, q^0<0\,,
    \end{cases} \quad \text{and fixed} \quad \mbs{b}\,,
\end{equation}
the integrand in eq.~\eqref{eq:scalarIntWaveform} is exponentially suppressed (see eq.~\eqref{eq:expSupp}) and thus vanishes. This condition fixes
\begin{equation}
    c_2=c_1\,,
\end{equation}
in eq.~\eqref{eq:newExBMIsUnfixed}.

To fix $c_1$, we evaluate eq.~\eqref{eq:scalarIntWaveform} at zero impact parameter. In doing so, it is convenient to commit to the rest frame of $u_1$, where
\begin{equation}
    u_1=(1,0,\mbs{0}_\perp) \quad \text{and} \quad u_2=(y,\sqrt{y^2{-}1},\mbs{0}_\perp)\,,
\end{equation}
such that $u_1{\cdot}u_2=-y$. In this frame, the integral at $b^\mu=0$ becomes 

\begin{equation}
    \begin{split}
\hspace{-0.25cm}\mathcal{I}_{\alpha}\vert_{b=0}&=\int_{\mathcal{M}}\dbar^\D  q \frac{\delta(q^0)\delta(yq^0{-}\sqrt{y^2{-}1}q^1{-}w_2)}{[-(q^0)^2+(q^1)^2+(\mbs{q}_\perp)^2-i\varepsilon]^{\alpha}
}
\,.
    \end{split}
\end{equation}
Integrating out the two delta functions gives the $\alpha^{\text{th}}$ mass derivative of the Euclidean $(\D{-}2)$-dimensional tadpole of mass $\mathfrak{m}=w_2/\sqrt{y^2-1}$

\begin{equation}
\begin{aligned}
\hspace{0cm}\mathcal{I}_{\alpha}\vert_{b=0}&=\frac{1}{2\pi\sqrt{y^2{-}1}}\partial^{(\alpha)}_{\mathfrak{m}^2}\int_{\mathbbm{R}^{\D-2}}\frac{\dbar^{\D-2}\mbs{q}_\perp}{(\mbs{q}_\perp)^2+\mathfrak{m}^2}\\&=\frac{\Gamma(\alpha{+}1{-}\D/2)}{2^{\D/2}\pi \Gamma(\alpha)\sqrt{y^2{-}1}}\left[\frac{y^2{-}1}{w_2^2}\right]^{\alpha{+}1{-}\D/2}\,.
\end{aligned}
\end{equation}

Using these results, and by computing the $b\to0$ limit of eq.~\eqref{eq:newExBMIsUnfixed}, $c_1$ is fixed to
\begin{equation}
c_1=\frac{1}{4} \csc \left(\frac{\pi  \D}{2}\right) \,.
\end{equation}

Putting everything together, the final expressions for the dimensionful master integrals read

\begin{subequations}\label{eq:wvfRes}
    \begin{align}
\mathcal{I}_1&=\frac{\left(b^2/w_2^2\right)^{\frac{4{-}\D}{4}}}{2 \pi \left(y^2{-}1\right)^{\frac{\D{-}2}{4}}}K_{\frac{4{-}\D}{2}}\left(\frac{\sqrt{b^2} w_2}{\sqrt{y^2{-}1}}\right)\,,\label{eq:I1sol}\\
\mathcal{I}_2&=\frac{\left(b^2/w_2^2\right)^{\frac{6{-}\D}{4}}}{4 \pi \left(y^2{-}1\right)^{\frac{\D{-}4}{4}}}K_{\frac{6{-}\D}{2}}\left(\frac{\sqrt{b^2} w_2}{\sqrt{y^2-1}}\right)\,.
    \end{align}
\end{subequations}

where $K_\nu(z)$ stands for the modified Bessel function of the second kind. The $\D\to4$ limit of eq.~\eqref{eq:I1sol} is smooth and agrees with \cite[eq.~ (C16)]{Cristofoli:2021vyo}, once convention differences are taken into account. The $\D$-dimensional analytic expressions in eq.~\eqref{eq:wvfRes} are new and constitute one of the main results of this work.

%----------------------------------
\subsection{QCD color dipole scattering}\makeatletter\def\@currentlabel{(C)}\makeatother\label{ssec:c}

\begin{figure}
    \centering
   \includegraphics[scale=1]{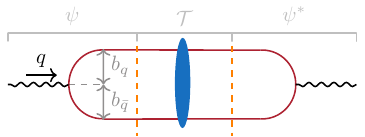}
    \caption{The bare color dipole cross-section discussed in the main text. The eikonal quark and anti-quark are represented by (red) Wilson lines. The color potential, which appears as a static two-dimensional pancake in the laboratory rest frame, models the highly boosted target nucleus and is represented by the blue region.}
    \label{fig:BKsetup} 
\end{figure}

A central objective of future electron-ion collision experiments \cite{Accardi:2012qut} is to gather data on how the density of partons inside hadrons changes as a function of energy. It is theorized that, as energy increases, this density becomes larger and larger until it reaches the so-called \emph{saturation regime} of QCD, where non-linear effects from gluon recombination ($gg\to g$) take over soft bremsstrahlung. This prediction arises in the \emph{color glass condensate} interpretation of \emph{deep inelastic scattering} (DIS) \cite{Iancu:2000hn,Ferreiro:2001qy}. In this framework, the incoming lepton emits a high-energy virtual photon scattering from the color potential of the proton. This interaction is then modeled in the frame where the virtual photon fluctuates into a color dipole (quarkonia) that scatters eikonally from the color potential (see Fig.~\ref{fig:BKsetup}). 

At leading order, the total cross-section for the photon polarization states is obtained by applying the optical theorem to the color dipole forward amplitude $\mathcal{T}$ \cite{Gelis:2012ri}
\begin{equation}
\begin{split}
    \hspace{-0.2cm}\sigma^{\gamma^\ast p}_{\text{LO}}=2\int \text{d}^2 \mbs{b}_q \text{d}^2 \mbs{b}_{\bar{q}} \text{d} z &\vert \psi(\Delta_{\perp},q^2,z)\vert^2 \mathcal{T}(\mbs{b}_q,\mbs{b}_{\bar{q}},Y)\,.
\end{split}
\end{equation} 

Here, $\psi=\psi_{\gamma^\ast\uparrow q\bar{q}}$ denotes the lightcone wavefunction of the virtual photon of momentum $q$ in the frame where it decays into a quarkonia dipole of transverse size $\Delta_{\perp}{=}\vert\mbs{b}_q{-}\mbs{b}_{\bar{q}}\vert$ carrying a fraction $z$ of the photon's longitudinal momentum. The forward amplitude $\mathcal{T}$ is related to the correlator of Wilson lines
\begin{equation}
    \tilde{U}(\mbs{b}_q,\mbs{b}_{\bar{q}},Y)=\frac{1}{N_\text{c}}\text{tr}[U(\mbs{b}_q,Y)U^\dag(\mbs{b}_{\bar{q}},Y)]\,,
\end{equation}
via $\mathcal{T}=1-\tilde{U}$. Here, $N_\text{c}$ denotes the number of colors and each Wilson line $U(\mbs{b}_\mathfrak{p},Y)$ represents a parton $\mathfrak{p}$ traversing the target at transverse position/impact parameter $\mbs{b}_\mathfrak{p}$ and rapidity $Y=Y(z)$ (see Fig.~\ref{fig:BKsetup}).

The rapidity evolution of the target color field is described by the Jalilian-Marian--Iancu--McLerran--Weigert--Leonidov--Kovner (JIMWLK) equation \cite{Mueller:2001uk}. An approximate, yet more tractable, large-$N_\text{c}$/mean-field description is given by the Balitsky--Kovchegov (BK) equation \cite{Balitsky:1995ub,Kovchegov:1999yj,BALITSKY_2001}, which is to leading order accuracy given by
\begin{equation}\label{eq:BKDEQ}
\begin{split}
    &\frac{\partial \tilde{U}(\mbs{b}_q,\mbs{b}_{\bar{q}},Y)}{\partial Y}=\int \text{d}^2\mbs{b}_{g} ~\mathcal{K}_{\text{BK}}^{\text{LO}}(\mbs{b}_q,\mbs{b}_{\bar{q}},\mbs{b}_{g})\\&\quad\times[\tilde{U}(\mbs{b}_q,\mbs{b}_{g},Y)\tilde{U}(\mbs{b}_{g},\mbs{b}_{\bar{q}},Y)-\tilde{U}(\mbs{b}_q,\mbs{b}_{\bar{q}},Y)]\,,
\end{split}
\end{equation}
where $\mathcal{K}_{\text{BK}}^{\text{LO}}(\mbs{b}_q,\mbs{b}_{\bar{q}},\mbs{b}_{g})=\frac{\alpha_\text{s} N_\text{c}}{2\pi^2}\frac{(\mbs{b}_{\bar{q}}-\mbs{b}_{q})^2}{(\mbs{b}_{\bar{q}}-\mbs{b}_{g})^2(\mbs{b}_{q}-\mbs{b}_{g})^2}$ and $\alpha_\text{s}$ is the strong coupling constant.\footnote{When considering high energy QCD in situations involving dilute targets and projectiles, the partonic Wilson lines in eq.~\eqref{eq:BKDEQ} tend to stay close to unity such that $\tilde{U}\to 1^-$ \cite{Caron-Huot:2016tzz}. In such scenarios, $\mathcal{T}$ is a small parameter and the relevant physics is governed by the linearized version eq.~\eqref{eq:BKDEQ} known as the 1-loop Balitsky--Fadin--Kuraev--Lipatov (BFKL) equation (see \cite{Fadin:1975cb,Balitsky:1978ic} and \cite{Gelis:2012ri} for a recent review).}

The solution to the BK equation predicts an interesting feature of the DIS total cross-section known as \emph{geometrical scaling} \cite{Iancu:2002tr}. This scaling is indicative of gluon saturation within the hadron in the Regge limit.

Over the past decade, significant efforts have been made to refine the BK equation by including next-to-leading order corrections and beyond (see, e.g., \cite{Balitsky:2007feb,Balitsky:2009xg,Caron-Huot:2015bja,Caron-Huot:2016tzz}). These refinements involve calculating higher-order corrections in the strong coupling constant, which can be quite cumbersome. In particular, as intermediate steps, it is often necessary to trade the transverse-momentum dependence in expressions in favor of transverse position. This step necessarily leads to complicated Fourier integrals.

\begin{figure}
    \centering
\includegraphics[scale=0.7]{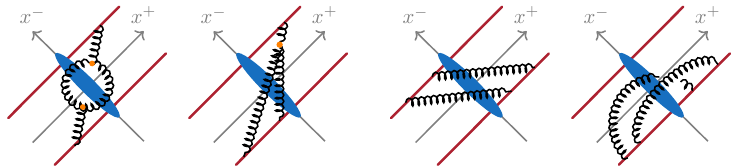}
    \caption{A small sample of NLO diagrams relevant to the rapidity evolution of a color dipole in lightcone coordinates (the transverse direction is left implicit). The first diagram exemplifies a cut self-energy correction, while the subsequent ones illustrate cut vertices. The Fourier integrals discussed in the main text emerge as intermediate steps in the computation of NLO BK observables in $\D$-dimensions.}
    \label{fig:BK} 
\end{figure}

As illustrative examples, we consider two $\D$-dimensional families of integrals relevant to deep inelastic scattering in the saturation regime
\begin{subequations}
    \begin{align}
    I^{ij} &= \int_{\mathbbm{R}^{2\D}} \dbar^{\D}q_{1}\dbar^{\D}q_{2}\frac{N_{I}^{ij}(q_{1},q_{2})\e^{i(q_{1}\cdot x_{1}{+}q_{2}\cdot x_{2})}}{q_{1}^{2}(q_{1}^{2}\tau{+}q_{2}^{2})}\,,\label{eq:BK1}\\
    G^{ij} &= \int_{\mathbbm{R}^{2\D}}\dbar^{\D}q_{1}\dbar^{\D}q_{2}\frac{N_{G}^{ij}(q_{1},q_{2})\e^{i(q_{1}\cdot x_{1}+q_{2}\cdot x_{2})}}{(q_{1}+q_{2})^{2}(q_{1}^{2}\tau+q_{2}^{2})}\,,\label{eq:BK2}
\end{align}
\end{subequations}
where the $q_i\equiv\mathbf{q}^{\perp}_i$s are Euclidean, $1\ge \tau> 0$ and
\begin{equation}
    \begin{split}
            N_{I}^{ij} &= q_{1}^{i}q_{2}^{j}\,,\\
            N_{G}^{ij} &= \delta^{ij}(q_{1}^{2}{-}q_{2}^{2}){-}\frac{2q_{1}^{i}(q_{1}{+}q_{2})^{j}}{u}{+}\frac{2(q_{1}{+}q_{2})^{i}q_{2}^{j}}{u\tau}\,.
    \end{split}
\end{equation}
In particular, in $\D=2$, \cref{eq:BK1,eq:BK2} appear in the derivation of the NLO BK equation \cite[eq.~(42)]{Balitsky:2007feb}. A small subset of diagrams leading to their appearance is shown in Fig.~\ref{fig:BK}.\\

In the following, we present new closed-form formulae for \cref{eq:BK1,eq:BK2} in $\D$ dimensions. We anticipate these results to be useful considering that the $\mathcal{O}(\epsilon)$ correction to the NLO BK equation yields non-trivial contributions to the NNLO BK equation in the critical dimension.\footnote{More precisely, as prescribed by the ``spacelike-timelike correspondence'' \cite{Weigert:2003mm,Hatta:2008st,Caron-Huot:2015bja}, at any fixed order in $\alpha_\text{s}$, the non-global log Hamiltonian is independent of $\epsilon$ in dimensional regularization and equals the BK Hamiltonian in the critical dimension (recall that non-global observables (e.g., jet shapes) involve incomplete/``non-global'' integrals over final states phase space. These phase-space cuts lead soft radiation to not be integrated over \emph{all} angles, resulting in ``non-global'' large logarithms that need to be resummed). Concretely,
\begin{equation*}
\text{\emph{if}}~~H_{\text{BK}}^{(2)}=H_{\text{BK}}^{(2,0)}{+}\epsilon H_{\text{BK}}^{(2,1)}{+}...~~\text{\emph{then}}~~H_{\text{BK}}^{(3)}{-}H_{\text{NGL}}^{(3)}=H_{\text{BK}}^{(2,1)}\,.
\end{equation*}
This situation bears similarity to the relation between the soft anomalous dimension $\gamma_\text{s}$, which is independent of $\epsilon$, and the rapidity anomalous dimension, as mentioned in \cite[eq.~(6.21)]{Vladimirov:2017ksc}.}
\myparagraph{Tensor decomposition}
We first perform the tensor decomposition of $I^{ij}$ and $G^{ij}$, namely
\begin{align}\label{eq:TD}
    I^{ij} = \sum_{a=1}^{5}I_{a}t_{a}^{ij} \quad\text{and}\quad G^{ij} = \sum_{a=1}^{5}G_{a}t_{a}^{ij}\,,
\end{align}
with basis
\begin{equation}
    \begin{aligned}
    t_{1}^{ij} = x_{1}^{i}x_{1}^{j}, \quad t_{2}^{ij} &= x_{1}^{i}x_{2}^{j}, \quad t_{3}^{ij} = x_{2}^{i}x_{1}^{j},\\
    t_{4}^{ij} = x_{2}^{i}x_{2}^{j}, &\quad t_{5}^{ij} = \delta^{ij}\,.
\end{aligned}
\end{equation}

Here, $I_{a}$ and $G_{a}$ are respectively given by
\begin{align}
    \{I_{a},G_{a}\} = \sum_{b=1}^{5}(T^{-1})_{ab}\{K_{a}^{(I)},K^{(G)}_{a}\}\,,\label{eq:tensorTrans}
\end{align}
where we have defined
\begin{equation}
     K_{a}^{(I)} = t_{a}^{ij}I^{ij}\,, \quad K_{a}^{(G)} = t_{a}^{ij}I^{ij}\,, \quad T_{ab} = t_{a}^{ij}t_{b}^{ij}\,.
\end{equation}

Therefore, in order to find the tensor decompsoition of these integrals, our first task is to compute the scalar integrals $K_{a}^{(I)}$ and $K_{a}^{(G)}$ for $a=1,\dots,5$.\\

\myparagraph{Change of variables}
For $K_{a}^{(I)}$, we make the change of variables
\begin{align}
    q_{1}=\frac{\kappa_{1}}{\sqrt{\tau}\abs{x_{2}}} \quad\text{and}\quad q_{2}=\frac{\kappa_{2}}{\abs{x_{2}}}\,,
\end{align}
while for $K_{a}^{(G)}$, we instead consider
\begin{align}
    q_{1} = \frac{\kappa_{1}-\sqrt{\tau}\kappa_{2}}{\sqrt{\tau}\abs{x_{2}-x_{1}}}\quad\text{and}\quad  q_{2}=\frac{\sqrt{\tau}\kappa_{1}+\kappa_{2}}{\abs{x_{2}-x_{1}}}\,.
\end{align}
From there, we define the dimensionless vectors
\begin{subequations}
    \begin{align}
    \xi_{1}=\frac{x_{1}}{\sqrt{\tau}\abs{x_{2}}}, &\quad \xi_{2}=\frac{x_{2}}{\abs{x_{2}}}\,,\\
    \zeta_{1} = \frac{x_{1}+\tau x_{2}}{\sqrt{\tau}\abs{x_{2}-x_{1}}}, &\quad \zeta_{2} = \frac{x_{2}-x_{1}}{\abs{x_{2}-x_{1}}}\,,
\end{align}
\end{subequations}
such that both integrals take the universal form
\begin{subequations}
    \begin{align}
    K_{a}^{(I)} &= (\sqrt{\tau}x_{2}^{2})^{2-\D}\mathcal{I}\left(\xi_{1},\xi_{2};\mathcal{N}_{a}^{(I)}\right)\,,\\
    K_{a}^{(G)} &= \frac{(1+\tau)^{\D-3}}{(\sqrt{\tau}(x_{2}-x_{1})^{2})^{\D-2}}\mathcal{I}\left(\zeta_{1},\zeta_{2};\mathcal{N}_{a}^{(G)}\right)\,,
\end{align}
\end{subequations}
where
\begin{align}
    \mathcal{I}\left(\eta_{1},\eta_{2};\mathcal{N}\right) = \int_{\mathbbm{R}^{2\D}}\dbar^{\D}\kappa_{1}\dbar^{\D}\kappa_{2}\ \frac{\mathcal{N}\e^{i(\kappa_{1}\cdot\eta_{1}+\kappa_{2}\cdot\eta_{2})}}{\kappa_{1}^{2}(\kappa_{1}^{2}+\kappa_{2}^{2})}\,,\label{eq:Fam}
\end{align}
and $\eta_2^2=1$. The numerators are respectively given by
\begin{subequations}
    \begin{align}
    \mathcal{N}_{a}^{(I)} &= t_{a}^{ij}N_{I}^{ij}\left(\frac{\kappa_{1}}{\sqrt{\tau}\abs{x_{2}}},\frac{\kappa_{2}}{\abs{x_{2}}}\right)\,,\\
    \mathcal{N}_{a}^{(G)} &= t_{a}^{ij}N_{G}^{ij}\left(\frac{\kappa_{1}-\sqrt{\tau}\kappa_{2}}{\sqrt{\tau}\abs{x_{2}-x_{1}}},\frac{\sqrt{\tau}\kappa_{1}+\kappa_{2}}{\abs{x_{2}-x_{1}}}\right)\,.
\end{align}
\end{subequations}\\
\myparagraph{Master integrals of $\mathcal{I}$} 
In Baikov representation, the family of integrals defined in eq.~\eqref{eq:Fam} reads
\begin{align}
    \mathcal{I}(\eta_{1},\eta_{2};\mathcal{N}) = \int_{C_{R}} \d^{7}\mbs{z}\frac{u(\mbs{z})f(\mbs{z})}{z_{1}z_{2}}\,,
\end{align}
where $z_{1}=\kappa_{1}^{2}$, $z_{2}=\kappa_{1}^{2}+\kappa_{2}^{2}$, $z_{3}=\kappa_{1}\cdot\kappa_{2}$, $z_{4}=\kappa_{1}\cdot\eta_{1}$, $z_{5}=\kappa_{1}\cdot\eta_{2}$, $z_{6}=\kappa_{2}\cdot\eta_{1}$, $z_{7}=\kappa_{2}\cdot\eta_{2}$, and only $z_1$ and $z_2$ can appear as denominators. The twist is given by
\begin{align}
    u(\mbs{z})=-\frac{\e^{i(z_{4}+z_{7})}(\eta_{1}^{2}-(\eta_{1}\cdot\eta_{2})^{2})^{\frac{3-\D}{2}}}{2^{\D}\pi^{5/2}\Gamma((\D{-}3)/2)\Gamma(\D/2{-}1)}B(\mbs{z})^{\frac{\D-5}{2}}\,,
\end{align}
where $B(\mbs{z})$ is given in the ancillary file
and $f(\mbs{z}(\kappa_{1},\kappa_{2},\eta_{1},\eta_{2}))=\mathcal{N}$. In this case, there are two master integrals
\begin{subequations}
    \begin{align}
    J_{1}(\eta_{1},\eta_{2}) &= \mathcal{I}(\eta_{1},\eta_{2};\kappa_{1}^{2})=\int_{C_{R}} \d^{7}\mbs{z}\frac{u(\mbs{z})}{z_{2}}\,,\\
    \frac{J_{2}(\eta_{1},\eta_{2})}{(\D-2)^{2}} &=\mathcal{I}(\eta_{1},\eta_{2};1)=\int_{C_{R}} \d^{7}\mbs{z}\frac{u(\mbs{z})}{z_{1}z_{2}}\,,
\end{align}
\end{subequations}
where the factor $(\D-2)^2$ is introduced for later convenience.

Using intersection decompositions, the differential equations for the basis $\mbs{J}=(J_{1},J_{2})^{T}$ read
\begin{equation}
\begin{aligned}
    \partial_{\eta_{1}^{2}}\mbs{J}&=\boldsymbol{\Omega}_{\eta_{1}^{2}}\cdot\mbs{J}, \quad \boldsymbol{\Omega}_{\eta_{1}^{2}} = \begin{pmatrix}
        \frac{1-\D}{1+\eta_{1}^{2}}&0\\
        \frac{(\D-2)(1+\eta_{1}^{2})}{4\eta_{1}^{2}}&\frac{2-\D}{2\eta_{1}^{2}}
    \end{pmatrix}\,,\\
    \partial_{\eta_{1}\cdot\eta_{2}}\mbs{J}&=\mbs{0}\,.
\end{aligned}
\end{equation}
The solutions to this system of differential equations is
\begin{subequations}\label{eq:unfixedSolMIsExC}
    \begin{align}
    \hspace{-0.3cm}J_{1}(\eta_{1}) &= J_{10}(1+\eta_{1}^{2})^{1-\D}\,,\\
    \hspace{-0.3cm}J_{2}(\eta_{1}) &= J_{20}(\eta_{1}^{2})^{\frac{2{-}\D}{2}}{+}\frac{J_{10}}{2}\mathcal{F}^{(I)}\,,
\end{align}
\end{subequations}
with $\mathcal{F}^{(I)}={}_{2}F_{1}\left(\frac{\D{-}2}{2},\D{-}2,\frac{\D}{2},-\eta_{1}^{2}\right)$. 

 To fix the constants $J_{10}$ and $J_{20}$, we examine the boundary conditions at $\eta_{1}=0$.  For $\mbs{J}$, the limits evaluate to
\begin{subequations}
\begin{align}
J_{1}(0) &=
2^{\D-2}\Gamma(\D-1)
\,,\\
J_{2}(0) &=
2^{\D-3}\Gamma(\D-1)
\,,
\end{align}
\end{subequations}
as shown in app.~\ref{sec:ExCBoundary}. Thus, since eq.~\eqref{eq:unfixedSolMIsExC} yields $J_1(0)=J_{10}$, we immediately see that
\begin{align}
J_{10} &= 2^{\D-2}\Gamma(\D-1)\,.
\end{align}

To fix $J_{20}$ we use the fact that $J_2(0)$ is finite. This necessarily requires $J_{20}=0$.\\

\myparagraph{Computing $I^{ij}$} For $I^{ij}$, we have that $\eta_{1}=\xi_{1}$ and $\eta_{2}=\xi_{2}$. Thus, the master integrals are
\begin{subequations}
    \begin{align}
    J_{1}^{(I)} &= 2^{\D-2}\Gamma(\D-1)\left(1+\frac{x_{1}^{2}}{\tau x_{2}^{2}}\right)^{1-\D}\,,\\
    J_{2}^{(I)} &= 2^{\D-3}\Gamma(\D-1)\mathcal{F}^{(I)}\,,
\end{align}
\end{subequations}
with $\mathcal{F}^{(I)} = {}_{2}F_{1}\left(\frac{\D-2}{2},\D-2,\frac{\D}{2},-\frac{x_{1}^{2}}{\tau x_{2}^{2}}\right)$. 

We can then decompose the $\mathcal{I}$ integrals in eq.~\eqref{eq:Fam} in terms of these master integrals to find
\begin{subequations}
    \begin{align}
    K^{(I)}_{1} &= \frac{\sqrt{\tau} x_{12}\mathcal{J}}{(\sqrt{\tau} x_{2}^{2})^{\D-1}}, \quad K^{(I)}_{2} = \frac{\mathcal{J}}{(\sqrt{\tau} x_{2}^{2})^{\D-2}}\,,\\ K^{(I)}_{3} &= \frac{\sqrt{\tau} x_{12}^{2}\mathcal{J}}{x_{1}^{2}(\sqrt{\tau} x_{2}^{2})^{\D-1}}, \quad K^{(I)}_{4} = \frac{x_{12}\mathcal{J}}{x_{1}^{2}(\sqrt{\tau} x_{2}^{2})^{\D-2}}\,,\\ K^{(I)}_{5} &= \frac{\sqrt{\tau} x_{12}\mathcal{J}}{x_{1}^{2}(\sqrt{\tau} x_{2}^{2})^{\D-1}}\,,
\end{align}
\end{subequations}
where $x_{12}=x_{1}\cdot x_{2}$ and $\mathcal{J} = \left(\frac{\tau x_{2}^{2}-x_{1}^{2}}{2\tau x_{2}^{2}}J_{1}^{(I)}-J_{2}^{(I)}\right)$.

Applying the transformation described earlier in eq.~\eqref{eq:tensorTrans}, we find
\begin{equation}
    I_{1} = I_{3} = I_{4} = I_{5} = 0 \quad \text{and} \quad I_{2} = \frac{\sqrt{\tau}\mathcal{J}}{x_{1}^{2}(\sqrt{\tau}x_{2}^{2})^{\D-1}}\,.
\end{equation}

The final answer is then
\begin{align}\label{eq:Ianswer}
    I^{ij} = \frac{\sqrt{\tau}\mathcal{J}x_{1}^{i}x_{2}^{j}}{x_{1}^{2}(\sqrt{\tau} x_{2}^{2})^{\D-1}}\,.
\end{align}
In $\D=2$, we find
\begin{align}\label{eq:ILO}
    I^{ij} = -\frac{x_{1}^{i}x_{2}^{j}}{x_{2}^{2}(x_{1}^{2}+\tau x_{2}^{2})}\,,
\end{align}
in agreement with the known result given in \cite[eq.~(42)]{Balitsky:2007feb}. (Subleading terms in the $\D=2-2\epsilon$ expansion are provided in app.~\ref{sec:exp} for `QCD practitioners' convenience.)\\

\myparagraph{Computing $G^{ij}$} For $G^{ij}$, we have instead that $\eta_{1}=\zeta_{1}$ and $\eta_{2}=\zeta_{2}$. Thus, the master integrals are
\begin{subequations}
    \begin{align}
    J_{1}^{(G)} &= 2^{\D-2}\Gamma(\D-1)\left(\frac{(1+\tau)(x_{1}^{2}+\tau x_{2}^{2})}{\tau(x_{2}-x_{1})^{2}}\right)^{1-\D}\,,\\
    J_{2}^{(G)} &= 2^{\D-3}\Gamma(\D-1)\mathcal{F}^{(G)}\,,
\end{align}
\end{subequations}
with $\mathcal{F}^{(G)} = {}_{2}F_{1}\left(\frac{\D-2}{2},\D-2,\frac{\D}{2},-\frac{(x_{1}+\tau x_{2})^{2}}{\tau(x_{2}-x_{1})^{2}}\right)$.

We again decompose our integrals and transform back to our tensor decomposition in eq.~\eqref{eq:tensorTrans} to find
\begin{equation}
    \begin{aligned}
            G_{1} &= \frac{(1+\tau)^{\D-1}\left(Y_{1}J_{1}^{(G)}+2\tau(x_{2}-x_{1})^{2}J_{2}^{(G)}\right)}{u\tau^{1+\frac{\D}{2}}(x_{1}+\tau x_{2})^{2}((x_{2}-x_{1})^{2})^{\D}}\,,\\
    G_{2} &= (\tau-1)G_{1}\,,\quad
    G_{3} = 0\,,\quad
    G_{4} = -\tau G_{1}\,,\\
    G_{5} &= -\frac{(1+\tau)^{\D-2}\left(Y_{2}J_{1}^{(G)}+Y_{3}J_{2}^{(G)}\right)}{\tau^{\frac{\D}{2}}(x_{1}+\tau x_{2})^{2}((x_{2}-x_{1})^{2})^{\D}}\,,
    \end{aligned}
\end{equation}
where
\begin{subequations}
    \begin{align}
    Y_{1} &= 4\tau x_{12}+(\tau-1)(\tau x_{2}^{2}-x_{1}^{2})\,,\\
    Y_{2} &= (1+\tau)^{2}((\tau-1)x_{1}^{2}x_{2}^{2}+(x_{1}^{2}-\tau x_{2}^{2})x_{12})\,,\\
    Y_{3} &= 2\tau(x_{2}-x_{1})^{2}((\tau-1)x_{12}+x_{1}^{2}-\tau x_{2}^{2}))\,.
\end{align}
\end{subequations}
The final result is then given by eq.~\eqref{eq:TD} and reads
\begin{align}\label{eq:Ganswer}
    G^{ij} = \sum_{a=1}^{5}G_{a}t_{a}^{ij}\,.
\end{align}
In $\D=2$, we find that
\begin{align}
    \hspace{-0.2cm} G^{ij} = \frac{2\tau(x_{1}{-}x_{2})^{i}x_{2}^{j}{+}2x_{1}^{i}(x_{1}{-}x_{2})^{j}{+}u\tau(x_{2}^{2}{-}x_{1}^{2})\delta^{ij}}{u\tau(x_{2}{-}x_{1})^{2}(x_{1}^{2}{+}\tau x_{2}^{2})}\,,
\end{align}
which is also in agreement with \cite{Balitsky:2007feb}. (Once again, subleading terms in the $\epsilon$-expansion are provided in app.~\ref{sec:exp}.)\\

Let us close this application by noting that for both \cref{eq:Ianswer,eq:Ganswer}, the $\epsilon$-expansions (see app.~\ref{sec:exp}) involve only polylogarithms with rational coefficients in the kinematics. This observation resonates with the fact that the known results for NNLO BK in $\mathcal{N}=4$ super Yang-Mills involve only (weight three) polylogarithms with similar coefficients \cite{Caron-Huot:2016tzz}. Therefore, one might naively expect the full QCD result to be constructed from similar functions.\\

The differential equation matrices evaluated in applications \ref{ssec:a}, \ref{ssec:b} and \ref{ssec:c} by means of intersection numbers, have also been verified by computing integration-by-parts identities in momentum space.

%----------------------------------
%----------------------------------
\section{\label{sec:conclusion}Conclusions}
In this letter, we applied intersection theory to dimensionally regularized Fourier integrals, extending the range of applicability of this technique in particle physics beyond Feynman integrals. 

We showed how to express Fourier integrals as twisted periods by deriving their Baikov representation. From there, we explained how to derive relations between Fourier integrals within a given family and how to obtain the system of differential equations they satisfy using intersection numbers. This offered a fresh perspective on Fourier calculus, inspired by recent advances in Feynman calculus.

We showcased our method by computing explicitly relevant instances of Fourier integrals appearing in the tree-level spectral waveform within pure general relativity and $\mathcal{N}=8$ supergravity, as well as in color dipole scattering at next-to-leading order (NLO) in QCD. Each time, we were able to verify our new (dimensionally regularized) results with existing data from the literature in specific limits, highlighting the method's accuracy and potential. We hope that these developments open up a number of new directions, some of which are outlined below. 

It would be interesting to see to what extent the method described in this letter proves efficient in tackling the remaining integrals relevant to the tree-level spectral waveform (i.e., those with $\beta_2 \neq 0$ in eq.~\eqref{eq:ExpTreeGrav0}), and more ambitiously, those pertinent to the one-loop spectral waveform (these come with an additional loop integration on top of the impact parameter Fourier transform). To the authors' knowledge, none of these has been analytically computed in $\D$ dimensions. 

Similarly, it would also be interesting to investigate the extent to which our approach can be adapted to address the remaining steps in calculating the full NLO evolution of color dipoles within dimensional regularization (with or without the incorporation of running coupling corrections \cite{Kovchegov:2006vj,Balitsky:2007feb}; the latter requiring the computation of Fourier transforms involving additional logarithms
\cite{Kovchegov:2006vj}). As stressed in the main text, such NLO results would contain valuable non-trivial information about higher-order corrections to the rapidity evolution equation in critical dimension, which is relevant for the phenomenology of high-energy hadronic and nuclear content. 

From the mathematical point of view, the study of the cohomology of Fourier integrals, poses interesting questions on the type of systems of differential equations they obey, as well as on the type of functions that are expected to appear in their solutions, possibly involving new types of transcendental leading singularities/maximal cuts. Finally, it would be interesting to explore the extent to which tropical geometry \cite{Panzer_2022,Arkani-Hamed:2022cqe,Borinsky:2023jdv,hillman2023subtraction,arkanihamed2023loop} 
could be used in the context of Fourier integrals, offering new directions for their analytic and numerical evaluation.

\acknowledgments We thank Simon Caron-Huot, Vsevolod Chestnov, Hjalte Frellesvig, Aidan Herderschee, Manoj K.
Mandal, Matteo Pegorin and Henrik Munch for useful discussions. 
We thank Ian Balitsky, Sergio Cacciatori, Simon Caron-Huot, Hjalte Frellesvig, Federico Gasparotto, Sebastian Mizera, and Franziska Porkert for comments on the manuscript. In particular, we thank Sebastian Mizera for pointing out to us that in cases where $u$ has essential singularities (as is manifestly the case at infinity in eq.~\eqref{eq:FourierBaikov} due to the exponential term), the theory of twisted cohomology changes \cite{Matsumoto1998-2, majima2000}. However, all the examples considered in this letter seem to indicate that the simplified algorithm for the intersection number in \cite{Brunello:2023rpq} remains valid. We thank the organizers and participants of the RADCOR23 conference, which triggered the development of this project. M.G. is grateful to the Institute for Advanced Study for its hospitality while part of this work was being completed.
M.G.’s work is supported by the National Science and Engineering Council of Canada (NSERC) and the Canada Research Chair program.
The work of S.S. received support from Gini's Foundation, in Padova.

\appendix
\section{Derivation of the Baikov representation for Fourier integrals}\label{sec:BaikovFourier}
Let us consider the generic Fourier integral given by eq.~\eqref{eq:genericFourier}. To derive the Baikov representation, we first split each momentum $q_{j}$ into components perpendicular and parallel to the subspace spanned by $\{q_{j+1},\dots,q_{L},x_{1},\dots,x_{E}\}$. The effect on the measure is 
\begin{align}\label{eq:measure0}
    \d^{\D}q_{j} = \d^{E-L-j}q_{j\parallel}\wedge\d^{\D-E+L+j}q_{j\bot}\,.
\end{align}
The parallel component gives
\begin{align}
    \d^{E-L-j}q_{j\parallel} = \frac{\bigwedge_{k=j+1}^L\d(q_{j}\cdot q_{k})\wedge \d(q_{j}\cdot y_{E})}{G^{1/2}(q_{j+1},\dots,q_{L},x_{1},\dots,x_{E})}\,,\label{eq:measure1}
\end{align}
while the perpendicular component gives (in spherical coordinates)
\begin{align}
            \d^{N}q_{j\bot} &= q_{j\bot}^{N-1}\d\Omega_{N-2}\wedge \d q_{j\bot}\notag \\
    &= \frac{1}{2}q_{j\bot}^{N-2}\d\Omega_{N-2}\wedge \d(q_{j\bot}^{2})\label{eq:measure2}\\
    &= \frac{1}{2}\left(\frac{G(q_{j},\dots,x_{E})}{G(q_{j+1},\dots,x_{E})}\right)^{\frac{N-2}{2}}\d\Omega_{N-2}\wedge\d(q_{j\bot}^{2})\,,\notag 
\end{align}

where $N=\D-E+L+j$ and $\d\Omega_{N-2}$ is the measure accounting for the $N-2$ angular degrees of freedom. Substituting \cref{eq:measure1,eq:measure2} into eq.~\eqref{eq:measure0}, we observe that the Gram determinants simplify, leading to 
\begin{align}
\int_{\mathbbm{R}^{L\D}}\prod_{j=1}^{L}\dbar^{\D}q_{j} &= \frac{\pi^{-\frac{L(L-1)}{4}-\frac{LE}{2}}G(x_{1},\dots,x_{E})^{\frac{-\D+E+1}{2}}}{2^{\frac{L\D}{2}}\prod_{j=1}^{L}\Gamma\left(\frac{\D-L-E+j}{2}\right)}\notag\\&\hspace{-0.5cm}\times\int_{\mathbbm{R}^{n}} G(q_{1},\dots,x_{E})^{\frac{\D-L-E-1}{2}}\prod_{j=1}^{n}\d S_{j}\,.
\end{align}
We then conveniently redefine our propagators such that
\begin{align}
    z_{i} = A_{ij}S_{j}+B_{j}\,,
\end{align}
which introduces a Jacobian for the change of variables, given by $\det A$. Thus, we find
\begin{align}
    \int_{\mathbbm{R}^{L\D}}\prod_{j=1}^{L}\dbar^{\D}q_{j} = \kappa \int_{\mathbbm{R}^{n}}\d^{n}\mbs{z}B(\mbs{z})^{(\D-L-E-1)/2}\,,
\end{align}
where $\kappa$ and $B(\mbs{z})$ are defined in \cref{eq:baikovconst,eq:baikovPoly}, respectively.\\
\section{Fourier transform of a power-law}\label{sec:fourierPL}
From the generalized Schwinger trick, we have
\begin{equation}
\begin{split}
        \int_{\mathbbm{R}^\D}\dbar^\D q\frac{\e^{i q \cdot x}}{A(q)^{n}} &= \frac{2\pi^{n/2}}{\Gamma(n/2)}\int_{\mathbbm{R}^\D}\dbar^\D q~\e^{i q \cdot x}\\&\quad \times \int_{0}^{\infty}\d\xi~\xi^{n-1} \e^{-\pi \xi^2 A(q)^2} \,. \label{eq:todo0}
\end{split}
\end{equation}
For $A(q)=q$, and after swapping the order of integration on the right-hand side, the inner integral becomes much easier to perform. This is because the Fourier transform of a Gau\ss ian is itself a Gau\ss ian
\begin{equation}
    \int_{\mathbbm{R}^\D}\dbar^\D q~\e^{-\pi \xi^2 q^2} \e^{i q \cdot x} = \frac{\xi^{-\D}\e^{-\frac{x^2}{4\pi\xi^2}}}{(2\pi)^{\D/2}}\,. 
\end{equation}
Plugging this into eq.~\eqref{eq:todo0}, we find (for $0<n<\D$)
\begin{equation}\label{known1}
    \begin{split}
\eqref{eq:todo0}\vert_{A=q}&=N\int_{0}^{\infty} \d\xi\xi^{n{-}\D{-}1}\e^{-\frac{x^2}{4\pi\xi^2}}  \\&= N\int_{0}^{\infty} \d\zeta\zeta^{(\D-n)-1} \e^{-\frac{x^2\zeta^2}{4\pi}}
\\&=\frac{\Gamma((\D-n)/2)}{2^{n-\D/2}\Gamma(n/2)} (x^2)^{\frac{n-\D}{2}}\,,         
    \end{split}
\end{equation}
where $N=\frac{2\pi^{n/2}}{(2\pi)^{\D/2}\Gamma(n/2)}$ and $\zeta = 1/\xi$. After a trivial change of variables, eq.~\eqref{known1} is seen to agree with \cite[eq.~(A.1)]{Caron-Huot:2022lffq}.
\section{Ultra-soft graviton spectral waveform}\label{sec:exBeasy}
As complementary material to the results presented in sec.~\ref{ssec:a}, we examine the integral in eq.~\eqref{eq:ExpTreeGrav0} in the ultra-soft graviton ($k^\mu=0$) regime 
\begin{align}
    I_{\alpha}^{\mu} = \int_{\mathcal{M}}\dbar^{\D} q\delta(u_{1}\cdot q)&\delta(u_{2}\cdot q)\frac{\e^{-iq\cdot b}q^{\mu}}{(q^{2})^{\alpha}}\,,\label{impulse}
\end{align}
using intersection theory and differential equations. This integral, while solvable through more conventional methods such as Schwinger parameters, remains a simple enough example to showcase the application of the techniques used to address the more challenging calculations discussed in the main text.\\

\myparagraph{Evaluation of $I_{\alpha}^{\mu}$} We have $L=1$ internal vector $\{q\}$ and $E=3$ external vectors $\{b,u_{1},u_{2}\}$, with kinematics 
$u_{1}^{2}=u_{2}^{2}=-1$, $u_{1}\cdot u_{2}=-y$, $u_{1}\cdot b=u_{2}\cdot b=0$. We define the $n=4$ integration variables such that $z_{1}=q^{2}$, $z_{2}=-iq\cdot b$, $z_{3}=u_{1}\cdot q$ and $z_{4}=u_{2}\cdot q$.

By performing a tensor decomposition, the integral takes the form
\begin{align}
    I_{\alpha}^{\mu} = I_{\alpha}^{(1)}b^{\mu}+I_{\alpha}^{(2)}u_{1}^{\mu}+I_{\alpha}^{(3)}u_{2}^{\mu}\,.
\end{align}
It is easy to see from eq.~\eqref{impulse} that contractions with $u_{1}$ or $u_{2}$ vanish due to the delta functions. Moreover, $u_{1}$ and $u_{2}$ are, by definition, transverse to the impact parameter $b$. Put together, these conditions give
\begin{align}
    0 = I_{\alpha}^{(2)}+y I_{\alpha}^{(3)} = y I_{\alpha}^{(2)}+I_{\alpha}^{(3)}\,.
\end{align}
and thus $I_{\alpha}^{(2)}=I_{\alpha}^{(3)}=0$. This implies
\begin{align}
    I_{\alpha}^{\mu} = I_{\alpha}^{(1)}b^{\mu} = \frac{b_{\nu}I_{\alpha}^{\nu}}{b^{2}}b^{\mu}\,, \label{eq:IaTensor}
\end{align}
such that only $I_{\alpha}^{(1)}$ needs to be computed. Its Baikov representation takes the form
\begin{equation}
        I_{\alpha}^{(1)}=\frac{1}{b^2}\int\d^{4}\mbs{z}\,\frac{i\,z_2}{z_{1}^{\alpha}}u(\mbs{z})\delta(z_{3})\delta(z_{4})\,,
\end{equation}
where the twist reads
\begin{align}
    &u(\mbs{z})=- i\frac{\left(-b^2 \left(y^2-1\right)\right)^{2-\frac{\D}{2}}}{2^{\D/2}\pi ^{3/2} \Gamma \left(\frac{\D-3}{2}\right)}\times \\
    &\hspace{-0.2cm}\left(b^2 \left(-\left(y^2-1\right) z_1-2 y z_4 z_3+z_3^2+z_4^2\right)-\left(y^2-1\right) z_2^2\right){}^{\frac{\D-5}{2}}\,.\notag
\end{align}
We can then integrate out $z_{3}$ and $z_{4}$ using the delta functions to find
\begin{align}
    I_{\alpha}^{(1)} = \frac{1}{b^{2}}\int\d z_{1}\d z_{2}\frac{i\,z_{2}}{z_{1}^{\alpha}}u(z_{1},z_{2},0,0)\,.
\end{align}
There is only one master integral ($\nu{=}1$), defined as
\begin{align}
    J = \int\dbar^{\D} q\delta(u_{1}\cdot q)\delta(u_{2}\cdot q)\frac{\e^{-iq\cdot b}}{q^{2}}\,.\label{eq:defJ}
\end{align}
$I_{\alpha}^{(1)}$ can then be decomposed onto this basis as
\begin{equation}
    I_{\alpha}^{(1)}=\frac{i(4-\D)}{b^2}J \,.
\end{equation}
The master integral $J$ obeys the differential equations
\begin{align}
    \partial_{b^{2}}J = \frac{4{-}\D}{2b^{2}}J \quad \text{and} \quad \partial_{y}J = \frac{-y}{y^{2}{-}1}J\,.\label{eq:JFromDE}
\end{align}

The solution to these equations is easily seen to be
\begin{align}
    J = J_{0}\frac{(b^{2})^{(4-\D)/2}}{\sqrt{y^{2}-1}}\,.
    \label{eq:JupToJ0}
\end{align}
To fix the boundary constant $J_{0}$ a direct evaluation of $J$ in eq.~\eqref{eq:defJ} is performed at $b^2=1$: first, the two delta functions in the rest frame of $u_2$ are integrated over. Up to a normalization factor, this gives an Euclidean $(\D{-}2)$-dimensional tadpole integral with numerator $\e^{-iq\cdot b}$. After mapping this result back to Schwinger parameter space, the integral can be easily evaluated for $b^2=1$, similarly to the computation outlined in app.~\ref{sec:fourierPL}. This allows us to fix

\begin{equation}
    J_0=-\frac{i2^{\frac{\D}{2}}}{32}\Gamma(\D/2-2)\,.\label{eq:J0}
\end{equation}
Putting everything together, we obtain
\begin{equation}
I_{\alpha}^{\mu}=i(4-\D) J_{0}\frac{(b^{2})^{(2-\D)/2}}{\sqrt{y^{2}-1}}q^{\mu}\,.
\end{equation}

\section{Boundary conditions for application \ref{ssec:c} }\label{sec:ExCBoundary}

We examine the $\eta_{1}=0$ limit of the master integrals in \ref{ssec:c}
\begin{align}
    J_{1}(\eta_{1}) = \mathcal{I}(\eta_{1},\eta_{2};\kappa_{1}^{2}), \quad J_{2}(\eta_{1}) = \mathcal{I}(\eta_{1},\eta_{2};1)\,,
\end{align}
with $\mathcal{I}(\eta_{1},\eta_{2},\mathcal{N})$ being the family of integrals defined in eq.~\eqref{eq:Fam}. If we set $\eta_{1}=0$, we have for the first master integral
\begin{align}
    J_{1}(0) 
    &= \int_{\mathbb{R}^{\D}}\dbar^{\D}\kappa_{2}\e^{i\kappa_{2}\cdot\eta_{2}}\int_{\mathbb{R}^{\D}}\frac{\dbar^{\D}\kappa_{1}}{\kappa_{1}^{2}+\kappa_{2}^{2}}\nonumber\\
    &= \frac{\Gamma\left(1-\frac{\D}{2}\right)}{2^{\frac{\D}{2}}}\int_{\mathbb{R}^{\D}}\dbar^{\D}\kappa_{2}\frac{\e^{i\kappa_{2}\cdot\eta_{2}}}{(\kappa_{2}^{2})^{1-\frac{\D}{2}}}\nonumber\\
    &= 2^{\D-2}\Gamma(\D-1)\,.
\end{align}
For the second master integral we have
\begin{align}
    &\frac{J_{2}(0)}{(\D-2)^{2}}=\int_{\mathbb{R}^{\D}}\dbar^{\D}\kappa_{2}\e^{i\kappa_{2}\cdot\eta_{2}}\int_{\mathbb{R}^{\D}}\frac{\dbar^{\D}\kappa_{1}}{\kappa_{1}^{2}(\kappa_{1}^{2}+\kappa_{2}^{2})}\nonumber\\
    &=\int_{\mathbb{R}^{\D}}\dbar^{\D}\kappa_{2}\e^{i\kappa_{2}\cdot\eta_{2}}\int_{0}^{1}\d x\int_{\mathbb{R}^{\D}}\frac{\dbar^{\D}\kappa_{1}}{(\kappa_{1}^{2}+\kappa_{2}^{2}x)^{2}}\nonumber\\
    &=\frac{\Gamma\left(2-\frac{\D}{2}\right)}{2^{\frac{\D}{2}}}\int_{\mathbb{R}^{\D}}\dbar^{\D}\kappa_{2}\frac{\e^{i\kappa_{2}\cdot\eta_{2}}}{(\kappa_{2}^{2})^{2-\frac{\D}{2}}}\times\left(\int_{0}^{1}\d x x^{\frac{\D}{2}-2}\right)\nonumber\\
    &=\frac{2^{\D-3}\Gamma(\D-1)}{(\D-2)^2}\,.
\end{align}
In order to find these boundary conditions, we have used the result from \cite[eq.(7.85)]{Peskin:1995ev}, the result from app.~\ref{sec:fourierPL}, Feynman parameters. 

\section{$\epsilon$-expansions for application \ref{ssec:c}}\label{sec:exp}
In this appendix, we provide explicit formulae for the $\epsilon$-expansions of \cref{eq:Ianswer,eq:Ganswer} up to $\mathcal{O}(\epsilon^2)$, namely
\begin{equation}
    \{I^{ij},G^{ij}\}=\sum_{n=0}^{2}\epsilon^n\big\{I^{ij}_{(n)},G^{ij}_{(n)} \big\}+\mathcal{O}(\epsilon^3)\,.
\end{equation}
The coefficients $I^{ij}_{(n)}$ are recorded explicitly in the text below, as well as in a computer-friendly form in the ancillary file \texttt{ancillary.m}. The coefficients $G^{ij}_{(n)}$ are much larger and, therefore, are only recorded in \texttt{ancillary.m}. 

For the former, setting $I^{ij}_{(n)}=x_1^i x_2^j I_{(n)}$, we have
\begin{subequations}
   \begin{align}
       I_{(0)}&=\eqref{eq:ILO}\,,\\
       I_{(1)}&=\frac{\Delta_- \log \left[\tfrac{\tau  x_2^2}{\Delta_+}\right]{-}2 x_1^2 \left(\log \left[\tfrac{\sqrt{\tau }
   x_2^2}{2}\right]{+}\gamma_\text{E} \right)}{\Delta_+ x_1^2 x_2^2}\,,\\
       I_{(2)}&=\Big[
       3 \Delta_- \log \left[\tfrac{\Delta_+}{\tau  x_2^2}\right] \left(\log \left[\tfrac{\Delta_+}{\tau  x_2^2}\right]{+}2
   \left(\log \left[\tfrac{\sqrt{\tau } x_2^2}{2}\right]{+}\gamma_\text{E} \right)\right)\notag\\&{+}3 \Delta_+ \text{Li}_2\left[\tfrac{-x_1^2}{\tau 
   x_2^2}\right]{+}6 x_1^2 \left(\log \left[\tfrac{\sqrt{\tau } x_2^2}{4}\right]{+}2 \gamma_\text{E} \right) \log \left[\sqrt{\tau }
   x_2^2\right]\notag\\&+x_1^2 \left(\pi ^2+6 (\gamma_\text{E}{-}\log 2)^2\right)\Big]/(3 \Delta_+ x_1^2 x_2^2)\,,
       %-\frac{}{3 \Delta_+ x_1^2 x_2^2}\,,
   \end{align} 
\end{subequations}
where $\Delta_\pm=x_1^2\pm \tau x_2^2$.

\bibliography{article}

\end{document}